\documentclass[showpacs,prl,reprint,superscriptaddress,preprintnumbers,twocolumn]{revtex4-2}
\usepackage{amssymb}
\usepackage{amsmath}
\usepackage{graphicx}
\usepackage{multirow}
\usepackage{hyperref}
\usepackage{xcolor}
\hypersetup{colorlinks,
	linkcolor={blue!50!black},
	citecolor={blue!50!black},
	urlcolor={blue!80!black}
}

\usepackage{color}

\begin{document}
\DeclareGraphicsExtensions{.pdf}

\title{Reply to \href{https://arxiv.org/abs/2103.10268}{arXiv:2103.10268} `Comment on ``Crossover of Charge Fluctuations across the Strange Metal Phase Diagram'''}
\author {Ali Husain}
\altaffiliation{Current address: University of British Columbia, Vancouver, BC V6T 1Z4, Canada}
\affiliation{Department of Physics and Materials Research Laboratory, University of Illinois at Urbana–Champaign, Urbana, IL 61801, USA}
\author {Matteo Mitrano}
\altaffiliation{Current address: Harvard University, Cambridge, MA 02138, USA}
\affiliation{Department of Physics and Materials Research Laboratory, University of Illinois at Urbana–Champaign, Urbana, IL 61801, USA}
\author {Melinda S. Rak}
\affiliation{Department of Physics and Materials Research Laboratory, University of Illinois at Urbana–Champaign, Urbana, IL 61801, USA}
\author {Samantha Rubeck}
\affiliation{Department of Physics and Materials Research Laboratory, University of Illinois at Urbana–Champaign, Urbana, IL 61801, USA}
\author {Bruno Uchoa}
\affiliation{Department of Physics and Astronomy, University of Oklahoma, Norman, OK 73069, USA}
\author {Katia March}
\affiliation{Department of Physics, Arizona State University, Tempe, AZ 85287, USA}
\author {Christian Dwyer}
\affiliation{Department of Physics, Arizona State University, Tempe, AZ 85287, USA}
\author {John Schneeloch}
\affiliation{Condensed Matter Physics and Materials Science Department, Brookhaven National Laboratory, Upton, NY 11973, USA}
\author {Ruidan Zhong}
\affiliation{Condensed Matter Physics and Materials Science Department, Brookhaven National Laboratory, Upton, NY 11973, USA}
\author {Genda D. Gu}
\affiliation{Condensed Matter Physics and Materials Science Department, Brookhaven National Laboratory, Upton, NY 11973, USA}
\author {Peter Abbamonte\footnote{abbamonte@mrl.illinois.edu}}
\affiliation{Department of Physics and Materials Research Laboratory, University of Illinois at Urbana–Champaign, Urbana, IL 61801, USA}

\date{\today}

\begin{abstract}

We recently reported \cite{mitrano18,husain19} measurements of the charge density fluctuations in the strange metal cuprate Bi$_{2.1}$Sr$_{1.9}$Ca$_{1.0}$Cu$_{2.0}$O$_{8+x}$ using both reflection M-EELS and transmission EELS with $\leq$10 meV energy resolution. We observed the well-known 1 eV plasmon in this material for momentum $q\lesssim$ 0.12 r.l.u., but found that it does not persist to large $q$. For $q\gtrsim0.12$ r.l.u., we observe a frequency-independent continuum, similar to that observed in early Raman scattering experiments \cite{bozovic1987,slakey1990}, that correlates highly with the strange metal phase \cite{husain19}.

In his Comment (\href{https://arxiv.org/abs/2103.10268}{arXiv:2103.10268}), J{\"o}rg Fink claims we do not see the plasmon, and that our results are inconsistent with optics, RIXS, and the author's own transmission EELS measurements with $\sim$100 meV resolution from the early 1990's \cite{nucker1989,nucker1991}. The author claims we have made a trigonometry error and are measuring a larger momentum than we think. The author asserts that the two-particle excitations of cuprate strange metals are accurately described by weakly interacting band theory in RPA with corrections for conduction band carrier lifetimes and Umklapp effects.

Here, we show that the author's Comment is in contradiction with established information from the literature. At $q\lesssim0.12$ r.l.u., we see the same 1 eV plasmon as other techniques. Moreover we compute our momentum correctly, adjusting the sample and detector angles during an energy scan to keep $q$ fixed. The only discrepancy is between our data and the results of Ref. \cite{nucker1989} for $q\gtrsim0.12$ r.l.u. where, because of the coarse resolution used, the data had to be corrected for interference from the elastic line. A reexamination of these corrections in early transmission EELS measurements would likely shed light on this discrepancy.

\end{abstract}

\maketitle
\section{Introduction}
We recently reported momentum-resolved electron energy-loss spectroscopy (M-EELS) measurements of the density fluctuation spectrum of the cuprate strange metal Bi$_{2.1}$Sr$_{1.9}$Ca$_{1.0}$Cu$_{2.0}$O$_{8+x}$ (Bi-2212) across its doping-temperature phase diagram \cite{husain19,mitrano18}. Using an energy resolution of 4 meV, we focused on momenta 0.1 $< q <$ 0.5 reciprocal lattice units (r.l.u., or units of $2\pi/a$ with $a=3.81 \textrm{ \AA}$) and energies from 0 to 2 eV. We found that the 1 eV plasmon peak seen in optics is clearly observable for $q\lesssim0.12$ r.l.u., as reported previously in Ref. \cite{vig17}. However, this excitation does not persist to large momenta. For $q\gtrsim0.12$ r.l.u., M-EELS shows a broad, featureless continuum of charge fluctuations extending up to 1 eV, which dominates the charge response over $\geq$90\% of the Brillouin zone. The detailed shape of this continuum depends on temperature and doping, but it is frequency-independent within the region of the phase diagram normally associated with strange metal \cite{proust16}. This continuum  resembles that observed in early Raman scattering measurements \cite{bozovic1987,slakey1990}, and bears a striking similarity to the marginal Fermi liquid (MFL) hypothesis that unites much of the basic phenomenology of strange metals \cite{varma89}. 

To test the validity of this result, we repeated these measurements using transmission EELS with 10 meV energy resolution in a Nion UltraSTEM microscope at Arizona State University \cite{husain19}. These experiments are momentum-integrating but, as a transmission measurement, directly reveal the bulk response. The results agreed quantitatively with the continuum observed in the M-EELS data, showing the same spectral shape and cutoff, establishing the continuum as a bulk effect \cite{husain19}.

In his Comment \cite{fink21}, J{\"o}rg Fink claims our results contradict optical \cite{levallois16}, resonant inelastic x-ray scattering (RIXS) \cite{nag2020}, and transmission EELS (T-EELS) \cite{nucker1989,nucker1991} studies. The author claims we compute our momentum incorrectly, failing to account for the energy change of the scattered electron, resulting in an incorrect value for ${\bf q} = {\bf k_i} - {\bf k_f}$, where $|{\bf k_f}| = \sqrt{2 m E_f}/\hbar$, during an energy-loss scan. 

In this Reply, we show that the plasmon measured with M-EELS is consistent with optics, RIXS and transmission EELS (T-EELS) measurements with $\sim$100 meV resolution in the low momentum regime, $q\lesssim0.12$ r.l.u. The M-EELS plasmon peak at $q=$ 0.05 r.l.u. has the same 1 eV energy and exhibits a similar lineshape to early T-EELS data at the nearest available momentum, $q=$ 0.06 r.l.u., as we reported previously \cite{vig17}. Further, we show that we compute our momentum correctly, using orientation matrix techniques developed for inelastic neutron scattering, in which the energy change of the scattered particle is properly treated by adjusting the sample and detector angles during an energy scan \cite{busing1967,vig17}. Use of a $UB$ matrix is the key difference between M-EELS and traditional reflection HR-EELS or ``R-EELS," which is carried out at fixed angles.

The only experimental discrepancy is between our measurements with 4 meV resolution and early T-EELS measurements with 100 meV resolution in the range $q\gtrsim0.12$ r.l.u., where T-EELS studies \cite{nucker1989} observe a plasmon we do not. In this regime, because of the coarse resolution used, the T-EELS data had to be corrected for interference from the elastic line. Quoting the original article, ``\textit{The loss function was derived from an EELS spectrum by subtracting contributions from the quasielastic peak and double scattering}" \cite{nucker1989}.  
A reexamination of these corrections would likely shed light on this discrepancy.

\section{Momentum accuracy of M-EELS}
We start by refuting the claim in Ref. \cite{fink21} that we determine our momentum incorrectly. 
Conventional high-resolution reflection EELS, usually referred to as HR-EELS \cite{ibachMillsBook}, measures the momentum-dependent charge density response of materials with high angular resolution \cite{ibachMillsBook}. As Ref. \cite{fink21} correctly points out, fixed angles do not result in fixed momentum transfer for low-energy electrons. Fixed-angle HR-EELS spectra trace out an \textit{arc} in momentum space whose extent depends on the range of energy loss (Fig. 1). 

\begin{figure}
	\includegraphics[width=8.2cm]{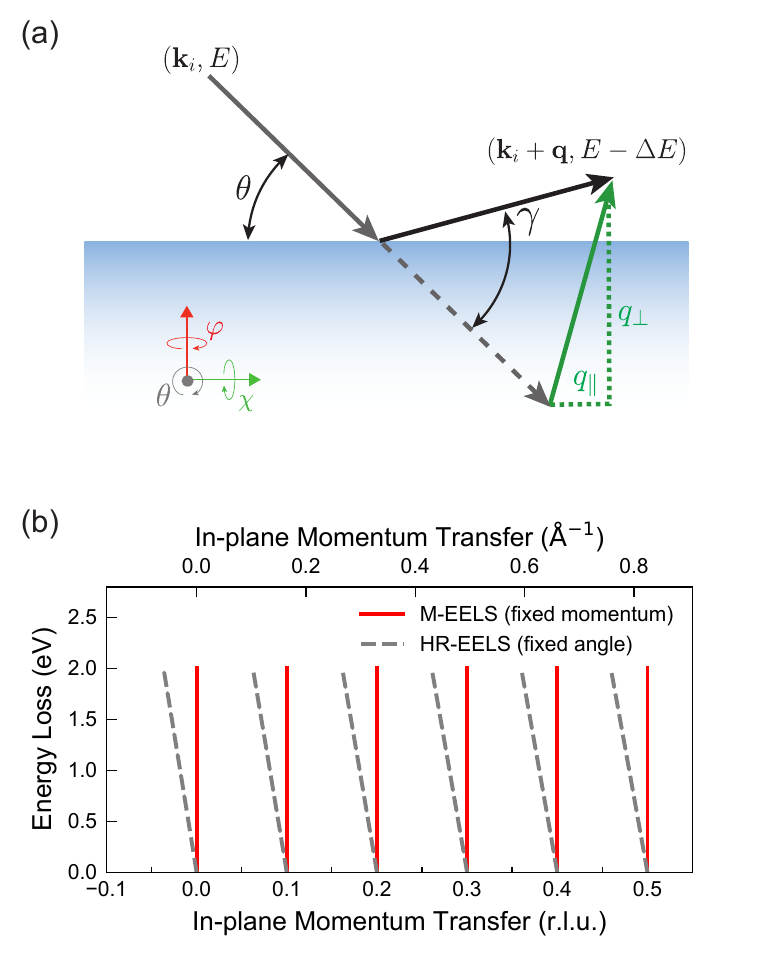}
	\caption{\textbf{(a)} Diagram of the scattering geometry of M-EELS showing the incident beam with momentum and energy $(\mathbf{k}_i,E)$ which is scattered and undergoes a momentum transfer $\mathbf{q}$ and energy loss $\Delta E$. The angle of the incident beam relative to the sample is $\theta$, and the angle between incident and scattered beams is $\gamma$. $\textbf{(b)}$ Comparison of $E\textrm{-}q$ cuts taken by M-EELS and HR-EELS calculated using the equations in the main text for a 50 eV incident beam energy and out-of-plane momentum transfer of 4.10 $\textrm{\AA}^{-1}$. Momentum is given in r.l.u. for Bi-2212 ($2\pi/a,$ with $a=3.81 \textrm{\AA}$). Notice that M-EELS acquires spectra at fixed momentum transfer for all energy losses, while the momentum transfer of HR-EELS depends on the energy loss and can vary by up 0.04 r.l.u. (0.07 $\textrm{\AA}^{-1}$) when $\Delta E=2$ eV.}
	\label{MEELS_vs_HREELS}
\end{figure}

This problem was originally encountered in inelastic scattering with thermal neutrons, where it was solved by use of the triple-axis spectrometer \cite{shirane2002}. In such a setup, using a eucentric sample goniometer, the detector and sample angles are varied simultaneously to maintain a fixed $\mathbf{q}$ as one scans the energy loss $\Delta E$. 

M-EELS was developed to solve the same problem for low-energy reflection HR-EELS \cite{vig17}. M-EELS employs a 5-axis eucentric sample goniometer and motorized analyzer, adjusting both the sample and detector angles during an energy scan, in concert with the lens voltages, to keep both the parallel and perpendicular components of the momentum transfer, $q_\parallel$ and $q_\perp$, fixed. This technique accounts for the energy change of the scattered electron in the same manner as a triple-axis instrument \cite{shirane2002}. The claim in Ref. \cite{fink21} that we do not do this is factually incorrect. 

In more detail, the momentum transfer parallel and perpendicular to a sample surface for reflection EELS of an electron with energy $E$ is given by 

\begin{eqnarray}
q_{\parallel} &=& k_f \cos(\gamma-\theta) - k_i \cos(\theta)\\
q_{\perp} &=& k_f \sin(\gamma-\theta) + k_i \sin(\theta).
\label{eqn:q}
\end{eqnarray}
Here, as shown in Fig \ref{MEELS_vs_HREELS}a, $q_{\parallel}$ is the momentum transfer parallel to the sample surface, $q_{\perp}$ is the momentum transfer perpendicular to the surface, $k_i=\sqrt{2m E/\hbar^2}$ is the incident electron momentum, $k_f=\sqrt{2m (E-\Delta E)/\hbar^2}$ is the final electron momentum, $\theta$ is the angle between the incident electron and sample surface, and $\gamma$ is the angle between the incident and scattered electrons. 

If the sample and scattering angles, $\theta$ and $\gamma$, are kept fixed, as done in HR-EELS, the momentum transfer will change as the energy loss $\Delta E$ is varied. M-EELS, like triple axis neutron spectrometers, resolves this issue by precisely varying the angles $(\theta, \gamma)$ in concert with the lens voltages to keep both $q_{\parallel}$ and $q_{\perp}$ fixed as $\Delta E$ is varied. The practical consequences of working at fixed momentum transfer, as opposed to fixed angles, are illustrated in Figure \ref{MEELS_vs_HREELS}b. 

In addition to correctly controlling the momentum, M-EELS employs techniques to properly register the {\bf q} vector with the crystal lattice of the material. This is accomplished by use of a $UB$ matrix, which is constructed by locating two noncolinear Bragg reflections from the sample \cite{busing1967}. $UB$ techniques are used in both neutron scattering and x-ray crystallography, where they allow accurate registry of the momentum with respect to the reciprocal lattice of the material. Use of a $UB$ matrix is what enabled our discovery of a soft plasmon in TiSe$_2$, which is the experimental signature of a Bose condensate of excitons \cite{kogar17}. 

\section{Consistency between M-EELS, Optics, and RIXS}
We now explain how to reconcile optical \cite{levallois16}, RIXS \cite{nag2020}, and M-EELS \cite{husain19,mitrano18} measurements of charge excitations in Bi-2212. Optical studies using ellipsometry and reflectivity demonstrate that the loss function of Bi-2212, at $q=0$, exhibits a plasmon peak around 1 eV with a full-width at half-maximum (FWHM) between $\sim$0.5 and 0.9 eV depending on the doping \cite{levallois16}. The lineshape of this plasmon is anomalous and shows non-Fermi liquid scaling for energies up to 0.6 eV in the strange metal regime \cite{marel2003}.

\begin{figure}
	\includegraphics[width=8.2cm]{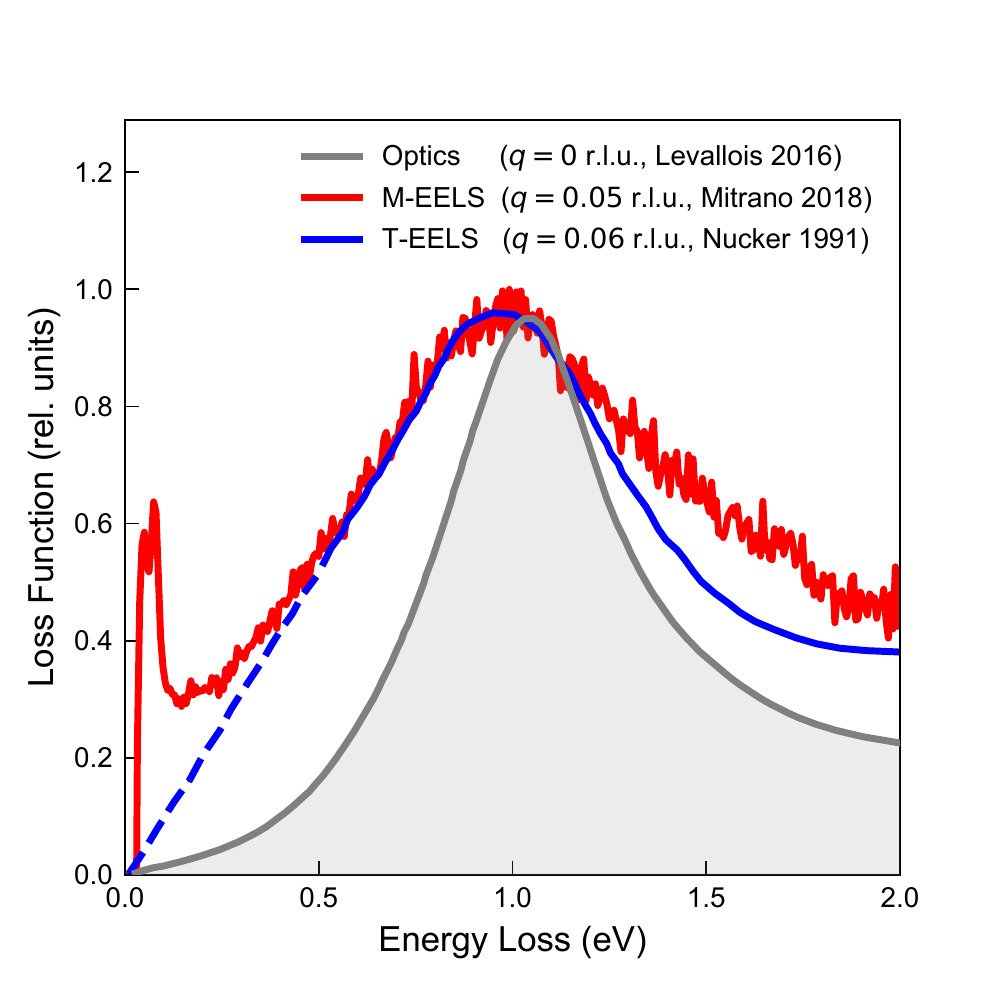}
	\caption{Comparison of the low-momentum charge response of optimally doped Bi-2212 at 300 K from ellipsometry (gray) at $q=0$ r.l.u. \cite{levallois16}, M-EELS (red) at $q=0.05$ r.l.u. \cite{mitrano18}, and T-EELS (blue) at $q=0.06$ r.l.u. \cite{nucker1991}. According to Ref. \cite{nucker1991}, the T-EELS data below 0.5 eV is extrapolated, so it is indicated by a dashed line. M-EELS and T-EELS agree reasonably well in this low-momentum regime and both see a broad plasmon peak. On the other hand, ellipsometry shows a significantly sharper plasmon peak, indicating that significant damping of the plasmon occurs even for $q \leq 0.05$ r.l.u.}
	\label{MEELS_TEELS_optics}
\end{figure}

More recently, RIXS studies of hole-doped cuprates have revealed dispersive, out-of-plane acoustic plasmon excitations in the low-momentum regime, $q<$ 0.15 r.l.u. \cite{nag2020,singh20}. In the case of Bi-2201, which is the closest proxy for Bi-2212 currently available from RIXS, the acoustic plasmon is highly damped and not clearly visible for $q>0.1$ r.l.u. \cite{nag2020}. Note that RIXS measurements have not yet observed the 1 eV plasmon seen in optics, which should be visible at out-of-plane momentum $L=2.0$ r.l.u. \cite{nag2020}, though such studies are making swift progress.

The claim in Ref. \cite{fink21} that there is a discrepancy between optics, RIXS, and M-EELS can be resolved by recognizing that these techniques probe different regions of momentum space. Optics measures $q\approx 0$, RIXS focuses on $0<q<0.1$ r.l.u., while M-EELS focuses on large momentum, $0.1<q<0.5$ r.l.u. At smaller momenta, $q\lesssim0.12$ r.l.u., M-EELS clearly observes the same 1 eV plasmon as other techniques (see Fig. \ref{MEELS_TEELS_optics} as well as Fig. 3 in Ref. \cite{vig17} and Fig. S1 of Ref. \cite{mitrano18}). Its energy and width are in good agreement with previous reflection and transmission EELS studies at small $q$ \cite{mitrano18,schulte2002,nucker1991}. As we explained in Appendix B of Ref. \cite{husain19}, the M-EELS continuum observed at $q\gtrsim0.12$ r.l.u. cannot persist all the way to zero momentum, since this would violate the compressibility sum rule \cite{mahan,nozieres99}. 

It is claimed in Ref. \cite{fink21} that the width of the acoustic plasmon in hole-doped cuprates observed with RIXS is about 0.1 eV. This statement is not consistent with the data. The FWHM of the plasmon in Bi-2201 at $L=1.75$ increases with momentum from $0.5\pm0.1$ eV at $q=0.05$ r.l.u., where the acoustic plasmon energy is 0.6 eV, to $0.76\pm0.07$ eV at $q=0.10$ r.l.u. where its energy is 0.84 eV (see Figs 4(a) and S12(e) of Ref. \cite{nag2020}). Because its width is comparable to its energy, it is appropriate to say that this acoustic plasmon is highly damped. Further, RIXS measurements of Bi-2212 at larger momenta, $q \sim 0.4$ r.l.u., observe broad isotropic charge fluctuations that are consistent with the continuum observed with M-EELS \cite{boschini2021}. Together, optics, RIXS, and M-EELS give a consistent picture that BSCCO exhibits a highly damped plasmon for small momenta, $q<0.12$ r.l.u., and an MFL-like continuum everywhere else in momentum space. 

\section{Comparison between M-EELS and Transmission EELS}
Momentum-dependent transmission EELS (T-EELS) studies with $\sim$100 meV resolution were performed on Bi-2212 in the 1980's and 1990s \cite{nucker1989,nucker1991} and claimed the charge response of Bi-2212 is entirely conventional, with an ordinary plasmon exhibiting $q^2$ dispersion from 1 eV at $q \sim 0$ up to 1.6 eV at $q=0.24$ r.l.u. At small momenta, these data are highly consistent with M-EELS. Fig. 2 compares the M-EELS spectrum at $q=0.05$ r.l.u. to the T-EELS spectrum at $q$ = 0.06 r.l.u. reproduced from Ref. \cite{nucker1991}, along with corresponding ellipsometry data at $q=0$ r.l.u. \cite{levallois16}. The M-EELS and T-EELS curves are in near quantitative agreement, showing the same energy and nearly the same lineshape (note that the T-EELS data were extrapolated to zero below 0.5 eV due to interference from the elastic line). This agreement was shown previously in Refs. \cite{vig17,mitrano18}. As pointed out in Ref. \cite{husain19}, the plasmon width observed with ellipsometry is about 50\% narrower than in both the M-EELS and T-EELS data, likely because of the small but nonzero momentum of the EELS measurements. 

At larger momentum, $q\gtrsim0.12$ r.l.u., there is a clear experimental discrepancy between M-EELS \cite{husain19,mitrano18} and early T-EELS data \cite{nucker1989,nucker1991}. The former, taken with an energy resolution of 4 meV, show a continuum that is energy-independent in the strange metal regime \cite{husain19}, while the latter, taken with $\sim$100 meV resolution, show a conventional Fermi liquid plasmon. This discrepancy needs to be explained. 

The likely explanation may be found in Ref. \cite{nucker1989}. Quoting the authors exactly, ``\textit{There is more background at higher momentum transfer due to the quasielastic line, which obscures the spectra below ~1 eV}.'' The intensity in T-EELS decreases rapidly with increasing $q$, so it becomes difficult at large momentum to resolve the loss spectrum from the background tail of the elastic line, which extends far beyond the nominal elastic FWHM of $\sim$100 meV. The spectra presented in Refs. \cite{nucker1989} and \cite{nucker1991} were corrected by subtracting both this elastic tail and double scattering features \cite{nucker1989}. The unaltered curves were not presented. The most likely explanation for the discrepancy lies in the details of how these subtractions were done.

\begin{figure}
	\includegraphics[width=8.2cm]{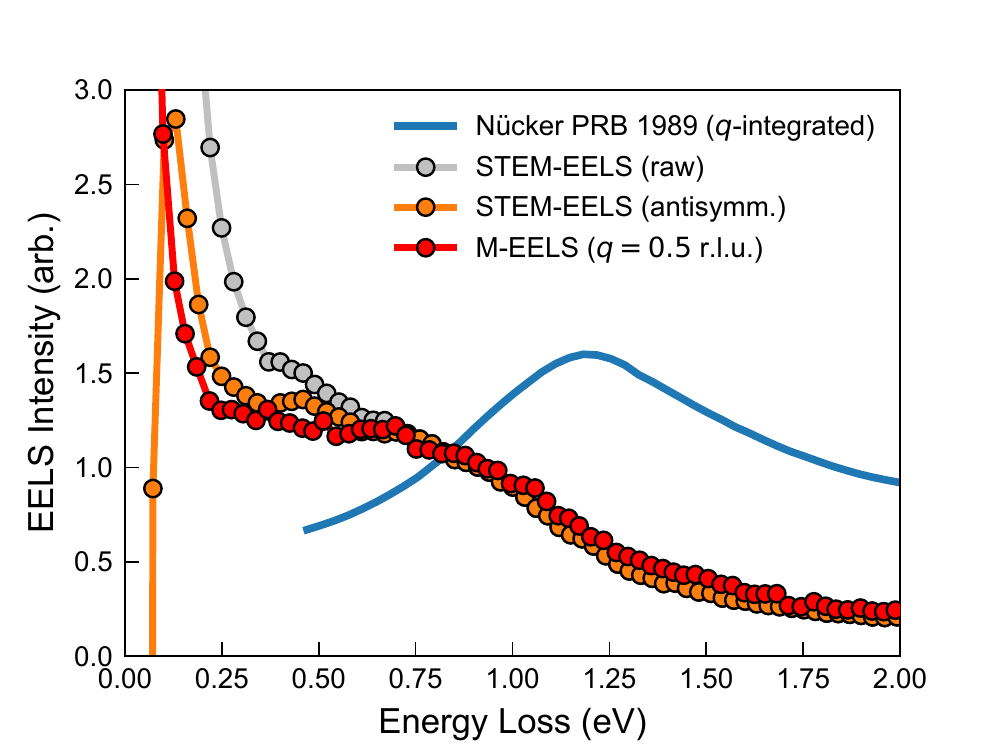}
	\caption{A comparison between momentum-integrated STEM-EELS data with 10 meV resolution (orange points) and an M-EELS measurement of the continuum at $q=0.5$ r.l.u. taken with 4 meV resolution (red points), reproduced from Ref. \cite{husain19}. Added to this plot is the momentum integral of the T-EELS data with $\sim$100 meV resolution from Ref. \cite{nucker1989} (blue line), showing marked disagreement with the other spectra.}
	\label{qinteg}
\end{figure}

To test the validity of our M-EELS data at $q\gtrsim0.12$ r.l.u., we replicated the experiments of Refs. \cite{nucker1989, nucker1991} by performing a transmission EELS measurement of optimally doped Bi-2212 using a Nion UltraSTEM microscope with 10 meV resolution at Arizona State University (published in Ref. \cite{husain19}). These measurements were done with a focused beam and so are momentum integrating. But they can be used to resolve the discrepancy between M-EELS and early T-EELS experiments. The results are shown in Fig. 3, which displays the STEM-EELS spectrum, an M-EELS measurement of the continuum at $q=0.5$ r.l.u. (reproduced from Ref. \cite{husain19}), and the momentum integral of the T-EELS curves in Ref. \cite{nucker1989}. The STEM-EELS and M-EELS data are nearly identical, showing the same continuum shape and cutoff (note that a nearly identical curve was independently measured in Ref. \cite{terauchi1999}). The T-EELS data, however, are inconsistent with the other curves. We conclude that the original  T-EELS data from Refs. \cite{nucker1989,nucker1991}, while accurate at low momentum (Fig. 2), are distorted at large $q$ by interference from the elastic line and subsequent data corrections. 

\section{Conclusion}
In summary, we have shown that the criticisms contained in Ref. \cite{fink21} are without merit, and are in conflict with established information in the literature.
Optics, RIXS, M-EELS and T-EELS measurements paint a consistent picture of a damped plasmon in hole-doped BSCCO for $q\lesssim0.12$ r.l.u. 

For $q\gtrsim0.12$ r.l.u., there is a discrepancy between modern instruments with $<$10 meV resolution early T-EELS measurements with 100 meV resolution, whose data had to be corrected for interference from the elastic line \cite{nucker1989,nucker1991}. The reason for the discrepancy most likely lies in the details of how these corrections were done. 

Considering only modern, high resolution data, all of which are available in open archives for public scrutiny, we conclude that for $q\gtrsim0.12$ r.l.u., which comprises  $\geq$90\% of the Brillouin zone, Bi-2212 exhibits a featureless continuum that is energy-independent throughout the strange metal region in the phase diagram \cite{proust16,husain19}. This continuum bears a striking similarity to the marginal Fermi liquid (MFL) hypothesis that unites much of the basic phenomenology of strange metals \cite{varma89}.

\bibliography{Reply_to_Fink}
\end{document}